\documentclass[12pt]{iopart}

\newcommand{\ord}{{\cal O}}
 
\newcommand{\ket}[1]{\left|#1\right>} 
\newcommand{\bra}[1]{\left<#1\right|} 
 
\newcommand{\bea}{\begin{eqnarray}} 
\newcommand{\ea}{\end{eqnarray}} 
 
\newcommand{\nn}{\nonumber\\}

\newcommand{\tdiff}{\frac{d}{dt}}

\begin{document}

\title[Decoherence versus Dynamical Casimir Effect]
{Decoherence versus Dynamical Casimir Effect} 

\author{Ralf Sch\"utzhold\footnote[7]{corresponding author:
{\tt schuetz@theory.phy.tu-dresden.de}}
and Markus Tiersch\footnote[6]{\tt markus@theory.phy.tu-dresden.de}}

\address{Institut f\"ur Theoretische Physik, 
Technische Universit\"at Dresden,\\
01062 Dresden, Germany}

\begin{abstract}
By means of two simple examples: phase and amplitude damping, the 
impact of decoherence on the dynamical Casimir effect is
investigated. 
Even without dissipating energy (i.e., pure phase damping), the
amount of created particles can be diminished significantly via the
coupling to the environment (reservoir theory) inducing decoherence. 
For a simple microscopic model, it is demonstrated that spontaneous
decays within the medium generate those problems -- Rabi oscillations
are far more advantageous in that respect.
These findings are particularly relevant in view of a recently
proposed experimental verification of the dynamical Casimir effect.
\end{abstract}


\section{Introduction}\label{Introduction}

Nearly a quarter of a century after the discovery of the static
Casimir effect \cite{casimir}, it has been realised that the
(non-inertial) motion of one of the mirrors should induce 
the creation of real particles out of the virtual quantum vacuum
fluctuations \cite{moore}.
Unfortunately, this striking prediction (the dynamical Casimir effect)
has not yet been verified experimentally.

To this end, it should be advantageous to exploit the phenomenon of
parametric resonance through a periodic perturbation of the discrete
eigen-frequencies of a finite cavity with the external (perturbation)
frequency matching twice the unperturbed eigen-frequency of one of the
cavity modes, for example.  
In the exact resonance case, the effective (time-averaged) Hamiltonian  
$\hat H_{\rm eff}$ can be derived by means of the rotating wave
approximation. 
If the aspect ratio of the cavity and the resonant mode is 
chosen such that there is no resonant inter-mode coupling 
(see, e.g., \cite{micha}), the effective Hamiltonian is just the
generator of a squeezing operator ($\hbar=c=1$ throughout)
\bea
\label{effH}
\hat H_{\rm eff}=i\,\xi\left[(\hat a^\dag)^2 - \hat a^2\right]
\,,
\ea
with $\xi$ depending on the strength of the perturbation etc., and   
$\hat a^\dag$, $\hat a$ denoting the usual bosonic creation and
annihilation operators for the resonant mode with the well-known
commutation relation $[\hat a,\hat a^\dag] = 1$. 
For example, if the perturbation is induced by an oscillation of the
dielectric permittivity of a thin dielectric slab with thickness
$a$ at one of the walls of the cavity with length $L \gg a$, the
squeezing parameter $\xi$ depends on the wave-number $k_\|$ parallel
to the thin slab and the external (perturbation) frequency $\omega$ in
the following way \cite{micha}
\bea
\label{xi}
\xi=\frac{1}{2}\,\frac{k_\|^2}{\omega}\,\frac{a}{L}\,\chi
\,,
\ea
with $\chi$ denoting the amplitude of oscillation of the inverse
dielectric permittivity. 
The dynamically squeezed vacuum state contains an exponentially
growing particle number 
\bea
\label{sinh}
\langle\hat n\rangle
=
\bra{0}\exp\{+i\hat H_{\rm eff}t\}
\,\hat n\,
\exp\{-i\hat H_{\rm eff}t\}\ket{0}
=
\sinh^2(2\xi t)
\,,
\ea
which should facilitate the measurement (in the ideal case).

Of course, aiming at an experimental verification of the dynamical
Casimir effect (cf.~\cite{exp}), deviations from the ideal behaviour
described above have to be taken into account as well. 
Previous studies have been devoted to the effects of detuning 
(i.e., a deviation of the external frequency from the exact resonance)
and losses (i.e., a finite quality factor of the cavity), 
see, e.g., \cite{losses}.
However, squeezed states such as $\exp\{-i\hat H_{\rm eff}t\}\ket{0}$
are highly non-classical states 
(see, e.g., \cite{squeezing,mandel,scully}) 
and therefore particularly vulnerable
to decoherence in a more general form (see, e.g., \cite{nielsen}).  

The impact of decoherence on the dynamical Casimir effect will be
studied in this Article by means of some representative examples for
decoherence channels as well as an explicit microscopic model.
It should be mentioned here that the following investigations are
devoted to the decoherence of the photon field due to the coupling to
the environment -- not to be confused with the decoherence of the
quantised position of a mirror due to the photon field 
(i.e, the dynamical Casimir effect) as discussed in \cite{via}.

\section{Reservoir}\label{Reservoir}

In the following, we discuss a very simplified model for the reservoir
-- which nevertheless yields some insight into the basic mechanism
and illustrates the main idea.
Let us consider a weak interaction of the mode $\hat a^\dagger,\hat a$
under investigation to a reservoir containing many degrees of freedom,
which will be approximated by independent two-level systems.
Without any loss of generality, the undisturbed Hamiltonian of the
two-level systems can be cast (after a suitable internal rotation)
into the form $\sum_I\omega_I\sigma_z^I$ with $2\omega_I$ denoting
their intrinsic energy gap and $\vec\sigma^I$ being the usual Pauli
spin matrices.  
The undisturbed Hamiltonian of the mode $\hat a^\dagger,\hat a$
without squeezing is just $\Omega\hat a^\dagger\hat a$.
In this simple model for the reservoir, the most general first-order
interaction Hamiltonian reads
\bea
\hat H_{\rm int}^{(1)}=
\sum\limits_I\left(
\hat a^\dagger\,\vec\lambda_I\cdot\vec\sigma_I+
\hat a\,\vec\lambda_I^*\cdot\vec\sigma_I\right)
\,,
\ea
where $\vec\lambda_I$ denote the small $|\vec\lambda_I|\ll\Omega$
first-order coupling constants.
Since the coupling is supposed to be small, we may apply the rotating
wave approximation after which only the resonant terms
($\omega_I=\Omega$) survive (see, e.g., \cite{scully,mandel})
\bea
\hat H_{\rm int}^{(1)}
\stackrel{\rm RWA}{=}
\sum\limits_{\omega_I=\Omega}
\lambda_I\left(\hat a^\dagger\sigma_-^I+\hat a\sigma_+^I\right)
\,.
\ea
Consequently, this effective Hamiltonian consists of hopping operators
and describes the decay channel.
Now let us consider the most general second-order interaction
Hamiltonian  
\bea
\hat H_{\rm int}^{(2)}
=
\sum\limits_I\left(
(\hat a^\dagger)^2\vec\zeta_I\cdot\vec\sigma_I+
\hat a^\dagger\hat a\,\vec\Lambda_I\cdot\vec\sigma_I+
\hat a^2\vec\zeta_I^*\cdot\vec\sigma_I
\right)
\,,
\ea
containing the second-order coupling constants $\vec\zeta_I$ and
$\vec\Lambda_I$ which are usually much smaller than the first-order 
couplings $\lambda_I$.
Again applying the rotating wave approximation
\bea
\hat H_{\rm int}^{(2)}
\stackrel{\rm RWA}{=}
\sum\limits_I\hat a^\dagger\hat a\,\Lambda_I\sigma_z+
\sum\limits_{\omega_I=2\Omega}
\zeta_I\left((\hat a^\dagger)^2\sigma_-^I+\hat a^2\sigma_+^I\right)
\,,
\ea
we observe that the second term describes resonant two-particle
hopping and is usually small compared to one-particle hopping 
$\hat H_{\rm int}^{(1)}$ since the second-order perturbation
$\zeta_I$ is supposed to be smaller than the first order
$\lambda_I$. 

However, the first term contains a sum over all reservoir modes --
without restrictions due to resonance conditions -- and thus may well
be of comparable magnitude or even larger than $\hat H_{\rm int}^{(1)}$.
This term corresponds to weak measurements (see, e.g., \cite{nielsen})
of the particle number $\hat a^\dagger\hat a$ without dissipating energy 
(as $[\hat a^\dagger\hat a\,\Lambda_I\sigma_z,\hat a^\dagger\hat a]=0$)
and is called phase damping.
In summary, we obtain the two leading contributions
\bea
\hat H_{\rm int}^{\rm RWA}
=
\sum\limits_{\omega_I=\Omega}
\lambda_I\left(\hat a^\dagger\sigma_-^I+\hat a\sigma_+^I\right)
+
\sum\limits_I\hat a^\dagger\hat a\,\Lambda_I\sigma_z
\,,
\ea
where the first (hopping) term describes amplitude damping (decay
channel) and the second contribution corresponds to phase damping.

After averaging over the degrees of freedom of the reservoir by means
of the usual (e.g., Born-Markov \cite{mandel,scully}) approximations,
we obtain the master equation in the Lindblad form 
(see, e.g., \cite{nielsen})
\bea
\label{Lindblad}
\frac{d\hat\varrho}{dt}=
-i\left[\hat H_{\rm eff},\hat\varrho\right]+
\sum\limits_\alpha
\left(2\hat L_\alpha\,\hat\varrho\,\hat L_\alpha^\dagger
-\left\{\hat L_\alpha^\dagger\hat L_\alpha,\hat\varrho\right\}
\right)
\,,
\ea
with two different Lindblad operators $\hat L_\alpha$.
The first one $\hat L_1$ arises from the first term in 
$\hat H_{\rm int}^{\rm RWA}$ and describes amplitude damping 
$\hat L_1=\sqrt{\gamma}\,\hat a$, whereas the second contribution in  
$\hat H_{\rm int}^{\rm RWA}$ generates phase damping with the Lindblad 
operator $\hat L_2=\sqrt{\Gamma}\,\hat a^\dagger\hat a$.
The two damping coefficients $\gamma$ and $\Gamma$ are determined by
the coupling constants $\lambda_I$ and $\Lambda_I$ as well as the
number of involved degrees of freedom of the reservoir.
Since this number is much larger for phase damping due to the absence
of resonance conditions, the phase damping rate $\Gamma$ can be as
large as or even larger than the amplitude damping rate $\gamma$ in
spite of the fact that the second-order coupling constants $\Lambda_I$
are usually much smaller than the first order $\lambda_I$.

\section{Master Equation and Particle Creation}

For the two decoherence channels discussed above, phase and amplitude
damping, we shall now investigate the impact of decoherence on the
amount of created particles $\hat n=\hat a^\dagger\hat a$. 
The mode under consideration $\hat a^\dagger,\hat a$ is described by
the effective density matrix $\hat \varrho$ whose evolution is
governed by the master equation (\ref{Lindblad}) 
\begin{equation}
\label{master}
\tdiff \hat\varrho = \xi [(\hat a^\dag)^2 - \hat a^2, \hat\varrho] +
\frac{\Gamma}{2} 
\left( 2\hat n \hat\varrho \hat n - \{ \hat n^2,\hat\varrho \} \right) +
\frac{\gamma}{2} \left( 2\hat a \hat\varrho \hat a^\dag - 
\{ \hat n,\hat\varrho \} \right)
\,,
\end{equation}
with the squeezing parameter~$\xi$ and the two damping
coefficients~$\gamma$ and $\Gamma$ representing the amplitude damping
rate and the phase damping rate, respectively. 

Instead of deriving the complete solution $\hat\varrho(t)$ of the
above equation, we just focus on the most interesting quantity, the 
time-dependent expectation value of the particle number
(in the interaction picture with $d\hat n/dt=0$)
\begin{equation}
\label{dn/dt-1}
\tdiff \langle \hat n \rangle
= 
\tdiff \Tr \left\{ \hat\varrho(t)\,\hat n\right\}
= 
\Tr\left\{ \left( \tdiff \hat\varrho(t) \right) \hat n \right\}
\,.
\end{equation}
After inserting the master equation (\ref{master}), only the first
term ($\xi$) and the amplitude damping term ($\gamma$) contribute 
(as phase damping does not dissipate energy), and, after some algebra, 
one obtains
\begin{equation}
\tdiff \langle \hat n \rangle
=
2\xi \langle \hat a^{\dag2} + \hat a^2 \rangle - 
\gamma \langle \hat n \rangle
\,.
\end{equation}
In view of the squeezing generated by $\hat H_{\rm eff}$, the
expectation value \mbox{$\langle \hat a^{\dag2} + \hat a^2 \rangle$}
will be non-zero in general. By using the same method as above, one
can derive the time-dependence of this quantity
\begin{eqnarray}
\tdiff \langle \hat a^{\dag2} + \hat a^2 \rangle
=
8\xi \langle \hat n \rangle -
(2\Gamma+\gamma) \langle \hat a^{\dag2} + \hat a^2 \rangle +4\xi
\end{eqnarray}
which fortunately closes the system of equations. 
This system of two first-order differential equations may be solved
easily and gives the general solution 
\begin{eqnarray}
\langle\hat n\rangle(t)
&=& 
C_+\, 
\frac{\Gamma + \sqrt{\Gamma^2 + 16\xi^2}}{8\xi}\,
\exp{ \left\{ -(\gamma+\Gamma) + \sqrt{\Gamma^2 + 16\xi^2}\right\} }   
\nn
&+& 
C_-\,
\frac{\Gamma - \sqrt{\Gamma^2 + 16\xi^2}}{8\xi}\, 
\exp{ \left\{ -(\gamma+\Gamma) - \sqrt{\Gamma^2 + 16\xi^2}\right\} }
\nn
&+&
\frac{8\xi^2}{2\Gamma\gamma + \gamma^2 -16\xi^2}
\,,
\end{eqnarray}
where $C_\pm$ are integration constants.
After incorporating the initial conditions 
${\langle\hat n\rangle(t=0)=n_0}$
and
$\langle (\hat a^\dag)^2 + \hat a^2 \rangle(t=0)=0$ 
(which holds for any initial density matrix that is diagonal in the
particle number basis such as a thermal ensemble) one arrives at
\begin{eqnarray}
\label{full}
\langle\hat n\rangle (t)
&=&
\frac{\sqrt{\Gamma^2+16\xi^2}+\Gamma}{4\sqrt{\Gamma^2+16\xi^2}} 
\left(2n_0 + \frac{\sqrt{\Gamma^2+16\xi^2}-\Gamma}
{\sqrt{\Gamma^2+16\xi^2}-\Gamma-\gamma} \right) 
e^{+(\sqrt{\Gamma^2+16\xi^2}-\Gamma-\gamma)t}  
\nn
&+&
\frac{\sqrt{\Gamma^2+16\xi^2}-\Gamma}{4\sqrt{\Gamma^2+16\xi^2}} 
\left(2n_0 + \frac{\sqrt{\Gamma^2+16\xi^2}+\Gamma}
{\sqrt{\Gamma^2+16\xi^2}+\Gamma+\gamma}\right) 
e^{-(\sqrt{\Gamma^2+16\xi^2}+\Gamma+\gamma)t}  
\nn
&-&
\frac{1}{2}\frac{16\xi^2}{\Gamma^2+16\xi^2-(\Gamma+\gamma)^2}
\,.
\end{eqnarray}
Note that the apparent divergence at 
$\sqrt{\Gamma^2+16\xi^2}=\Gamma+\gamma$
disappears since the singularities in the first and last terms cancel 
each other. 

As a consistency check, we reproduce Eq.~(\ref{sinh}) for the initial
vacuum state $n_0=0$ and no decoherence $\Gamma=\gamma=0$.
Furthermore, for negligible phase damping $\Gamma \ll \xi$, the
characteristic exponent is given by $4\xi-\gamma$ in accordance with 
\cite{losses}.
I.e., one only obtains exponential growth if the particle creation
process (i.e., squeezing) is faster than the decay $4\xi>\gamma$ 
(as one might expect, cf.~\cite{losses}). 
In the presence of additional phase damping, this threshold is shifted
to ${4\xi>\sqrt{\gamma^2+2\gamma\Gamma}}$ according to Eq.~(\ref{full}).

Let us study the behaviour at very short and late times,
respectively. 
For a positive amplitude damping rate $\gamma>0$, the particle number
$\langle\hat n\rangle (t)$ first decreases linearly in time  
(unless $n_0=0$, of course) as the squeezing mechanism sets in at
quadratic order $t^2$ only. 
At late times, the particle number either increases exponentially 
(above the threshold ${4\xi>\sqrt{\gamma^2+2\gamma\Gamma}}$) or
settles down to a constant (and for $\xi>0$ positive) value 
(below the threshold ${4\xi<\sqrt{\gamma^2+2\gamma\Gamma}}$).

\section{Decoherence without Dissipation -- Phase Damping}

Without amplitude damping $\gamma=0$ (high quality factor), 
the situation is qualitatively different from the scenario with 
dissipation.
For short times, the number of particles increases quadratically and
for late times, the particle number is always growing exponentially,  
but the characteristic exponent is reduced.
For fast decoherence rates $\Gamma\gg\xi$ (and a high quality factor),
the leading term behaves as  
\begin{eqnarray}
\label{fast}
\langle \hat n \rangle (t)
\approx
\left( n_0 + \frac{1}{2} \right)\,
\exp\left\{\frac{8\xi^2}{\Gamma}\,t\right\}
-\frac{1}{2}
\,,
\end{eqnarray}
i.e., the particle creation rate is strongly reduced by a factor of
order $\Gamma/\xi\gg1$. 
(Note that, for fast decoherence $\Gamma\gg\xi$ and intermediate times 
$1/\Gamma \ll t \ll \Gamma/\xi^2$, one obtains quasi-linear behaviour
according to the equation above.)

The reduction of the number of created particles can be understood in
the following way:
The permanent weak measurement of the particle number implies a
suppression of the off-diagonal elements of the density matrix 
(i.e., the interference terms) in the particle number basis 
(see, e.g., \cite{nielsen}). 
E.g., without squeezing $\xi=0$, we would have
\bea
\varrho_{mn}(t)
=
\bra{n}\hat\varrho(t)\ket{m}
=
\varrho_{mn}(0)\exp\left\{-\frac{\Gamma}{2}(m-n)^2t\right\}
\,.
\ea
A strong -- i.e., decisive -- measurement of the particle number would
set all off-diagonal elements to zero.
On the other hand, squeezing is a highly non-classical effect 
(see, e.g., \cite{squeezing,mandel,scully}) 
and the related mechanism for particle creation involves the transfer
of occupation from the diagonal to the off-diagonal elements and back
\bea
\varrho_{mn}=
\left(
\begin{array}{cccc}
\varrho_{00} & \rightarrow & \varrho_{02} & \dots \\
\downarrow & \varrho_{11} & \downarrow & \dots \\
\varrho_{20} & \rightarrow & \varrho_{22} & \dots \\
\vdots & \vdots & \vdots & \ddots
\end{array}
\right)
\,.
\ea
Therefore, the suppression of the off-diagonal elements diminishes
this effect. 
Roughly speaking, we lose the phase information/phase control which
is necessary for constructive interference and resonance.  

For fast decoherence $\Gamma\gg\xi$, one could visualise the major
effect via a {\em gedanken} experiment combining the time-evolution
governed by $\hat H_{\rm eff}$ with permanently repeated strong
measurements of the particle number at equidistant time intervals
$\Delta t=\ord(1/\Gamma)$. 
In the Heisenberg picture, the unitary evolution between two
measurements can be described by a Bogoliubov transformation 
\bea
\hat a(t+\Delta t)
=
\alpha(\Delta t)\hat a(t)+\beta(\Delta t)\hat a^\dagger(t)
\,,
\ea
with the Bogoliubov coefficients obeying the bosonic unitarity
relation $|\alpha|^2-|\beta|^2=1$.
For an initial state $\hat\varrho$ which is diagonal in particle
number basis (such as after a strong measurement of $\hat n$), the
expectation value evolves as
\bea
\langle\hat n(t+\Delta t)\rangle
=
\langle\hat n(t)\rangle+|\beta(\Delta t)|^2[1+2\langle\hat n(t)\rangle]
\,.
\ea
In the second term on the right-hand side, one may distinguish the
pure quantum vacuum effect $|\beta(\Delta t)|^2$ and the contribution
of classical resonance $2|\beta(\Delta t)|^2\langle\hat n(t)\rangle$.
For fast decoherence $\Gamma\gg\xi$, the time intervals are small 
$\xi\Delta t\ll1$ facilitating the introduction of an approximate
differential equation $\Delta t \to dt$ 
\bea
\tdiff \,\langle\hat n(t)\rangle
\propto
\frac{\xi^2}{\Gamma}[1+2\langle\hat n(t)\rangle]
\,,
\ea
where we have used 
$|\beta(\Delta t)|^2=\sinh^2(2\xi\Delta t)\approx4\xi^2\Delta t^2$ 
and $\Delta t=\ord(1/\Gamma)$.
We observe that the pure quantum vacuum effect 
$d\langle\hat n(t)\rangle/dt\propto\xi^2/\Gamma$ 
would only yield a linear increase $\propto t\xi^2/\Gamma$
(permanent reshuffling from diagonal to off-diagonal elements of 
$\varrho_{mn}$ and accumulating higher diagonal terms), 
but the classical resonance contributions 
$d\langle\hat n(t)\rangle/dt\propto2\langle\hat n(t)\rangle\xi^2/\Gamma$
amplify all particles present and thereby lead to an exponential
growth with a reduced exponent $\propto t\xi^2/\Gamma$ as in
Eq.~(\ref{fast}). 

\section{Microscopic Model}

In order to apply the above results to an explicit scenario, one has
to estimate the magnitude of the involved quantities -- such as the
phase and amplitude damping rates $\Gamma$ and $\gamma$,
respectively. 
This can be achieved either by experimental means, i.e., measuring the 
quantum coherence time ($\propto1/\Gamma$) and the quality factor 
($\propto1/\gamma$) of the used cavity etc., or via theoretical
calculations -- based on a microscopic model, for example.
In the following, we shall present a very simple microscopic model for
a material with a time-dependent index of refraction -- which,
nevertheless, yields some interesting conclusions. 

Let us consider a specifically designed semi-conductor at low
temperatures, for example, whose localised single valence electrons 
(far below the conducting band) can occupy three levels described by
the amplitudes $\psi_{a,b,c}(t)$. 
All the other levels are supposed to be unimportant and are not
occupied at all.
Illuminating the semi-conductor with a strong external Laser beam
tuned to the frequency of transition from the lowest-lying electronic
state $a$ to the first excited state $b$, we can manipulate the 
amplitudes of the electronic states $\psi_{a,b}(t)$.
In addition, a small and slowly varying electric test field ${\cal E}$ 
(e.g., micro-waves) is acting on the three-level system.
The spatial distribution of the ground state $a$ is supposed to be
very compact such that it is basically unaffected by the electric test
field ${\cal E}(t)$; but the excited states $b$ and 
$c$ are more spread out and hence couple to ${\cal E}(t)$ with the 
dipole moment $\kappa$ (dipole approximation). 
In the rotating wave approximation, the Lagrangian of the described
three-level system reads 
\bea
\!\!\!\!\!\!\!\!\!\!\!\!\!\!\!\!\!\!\!\!\!\!\!\!
{\cal L}
=
i\psi_a^*\dot\psi_a+i\psi_b^*\dot\psi_b+i\psi_c^*\dot\psi_c
-\Delta\omega\,\psi_c^*\psi_c
+
(\kappa\,{\cal E}(t)\psi_b^*\psi_c+\Omega_{\rm R}(t)\psi_a^*\psi_b+{\rm h.c.})
\,,
\ea
with $\Delta\omega$ denoting the energy gap between states $b$ and $c$, 
and $\Omega_{\rm R}(t)$ the Rabi frequency of the strong external
Laser field (see, e.g., \cite{mandel,scully}). 
Since ${\cal E}(t)$ is small, we may use linear response theory 
($\psi_c\ll\psi_a,\psi_b$), and employing the adiabatic
approximation\footnote{Outside the region of validity of the adiabatic
  approximation, $\Delta\omega$ should be replaced by the detuning
  between the frequency of the test field ${\cal E}$ and the
  $b\leftrightarrow c$ transition frequency $\Delta\omega$.}  
\bea
\kappa\,{\cal E}(t)\psi_b=
\Delta\omega\,\psi_c-i\dot\psi_c\approx\Delta\omega\,\psi_c
\,,
\ea
because ${\cal E}(t)$ is supposed to be slowly varying, we finally
(after eliminating $\psi_{c}$) arrive at the contribution to
the effective Lagrangian for the test field ${\cal E}$ 
\bea
\label{Leff}
{\cal L}_{\rm eff}
=
|\psi_b(t)|^2\frac{\kappa^2}{\Delta\omega}\,{\cal E}^2(t)
=
\frac{\varepsilon_{\rm eff}(t)-1}{2}\,{\cal E}^2(t)
\,.
\ea
Consequently, by means of a Laser-induced excitation of the
polarisable $b$-level $\psi_b(t)$, we may generate a time-dependent
effective dielectric permittivity $\varepsilon_{\rm eff}$.
This, in turn, can serve as a perturbation for the dynamical Casimir
effect \cite{micha,exp}.

For a rough estimate of the maximum possible order of magnitude of 
$\varepsilon_{\rm eff}$, we assume one three-level system per lattice
site with a dipole coupling $\kappa$ corresponding to the lattice
spacing of a few {\AA}ngstr\"oms (i.e., the electronic states $b$ and
$c$ are spread out over a few {\AA}ngstr\"oms).
The Rabi frequency $\Omega_{\rm R}$ determines the time-scale for the
dynamics of the micro-wave test field ${\cal E}$ and should be around 
a few GHz in order to achieve resonance \cite{exp}.
The adiabatic approximation employed above requires the energy gap 
$\Delta\omega$ to be much larger than that frequency.
If we tune the three-level system in order to make this energy gap 
as small as possible within the region of validity of the adiabatic
approximation, say $\ord(100\;\rm GHz)$, the effective dielectric
permittivity $\varepsilon_{\rm eff}$ can be as large as 
$\varepsilon_{\rm eff}\leq\ord(10^5)$, 
which is more than sufficient. 

\section{Spontaneous decay $\to$ noise/decoherence}

So far, we only considered stimulated transitions $a \leftrightarrow b$  
induced by the (external) Laser field.
Under this assumption, one obtains an effective Lagrangian (\ref{Leff}) 
leading to a unitary evolution, i.e., no decoherence.
As it will become evident below, this result changes drastically for a 
spontaneous decay back to the ground state $a$ instead of a stimulated
transition. 

Assuming that the spontaneous decay (involving optical frequencies) 
occurs much faster than the (slow) dynamics 
(e.g., micro-wave frequencies) of the test field ${\cal E}(t)$, we may 
adopt the sudden approximation and omit the time-dependence of 
${\cal E}(t)$ during that process.
The influence of the field ${\cal E}$ perturbs the polarisable
$b$-state inducing a mixing with the $c$-state and thus an
energy-shift, which can be calculated using second-order stationary
perturbation theory 
\bea
\label{stat}
E_b^{(2)}
=
E_b^{(0)}
+
\bra{b}\hat{\cal E}\ket{b}
+
\frac{|\bra{b}\hat{\cal E}\ket{c}|^2}{E_b^{(0)}-E_c^{(0)}}
\,.
\ea
The first-order term vanishes in the dipole approximation 
(cf.~the selection rules) $\bra{b}\hat{\cal E}\ket{b}=0$.  
Depending on the slowly varying electric field ${\cal E}(t)$, the
spontaneous decay of the excited state -- which is a mixture of the
states $b$ and $c$ -- releases a different energy 
$E_b^{(2)}[{\cal E}(t)]$ than the originally absorbed Laser photon in
general.   
Note that the totalised energy shift $E_b^{(2)}-E_b^{(0)}$ in
Eq.~(\ref{stat}) exactly corresponds to the polarisation term in
Eq.~(\ref{Leff}), i.e., the polarisation energy ${\cal E}{\cal P}$ of
the medium   
\bea
\label{polar}
\varepsilon_{\rm eff}{\cal E}^2
={\cal E}{\cal D}={\cal E}^2+{\cal E}{\cal P}
\,.
\ea
So far, the electric field ${\cal E}$ has been treated as a
classical perturbation in Eq.~(\ref{stat}) which is strictly valid in
the classical limit (coherent state with many micro-wave photons) only. 
For the quantised micro-wave radiation field in a dynamically squeezed
state, the situation is more complicated. 
Expressed in terms of the relevant mode $\hat a^\dag$, $\hat a$ with
the frequency $\Omega$, the electric field operator can be estimated
by 
\bea
\label{volume}
\hat{\cal E}^2=\ord\left(\frac{\Omega}{V}\,[\hat a^\dag+\hat a]^2\right)
\,,
\ea
with $V$ denoting the volume of the cavity.
In a squeezed state, the expectation value of $\hat{\cal E}^2$ and its
variance are strongly time-dependent in view of 
$\hat a^\dag(t)=e^{i\Omega t}\hat a^\dag(0)$. 
Since we may decompose the squeezed state into a linear superposition
of (nearly classical) coherent states, the energy of the emitted
photons will fluctuate randomly (depending on the moment of decay). 
Therefore, spontaneous decays inevitably lead to decoherence --
first, by generating energy noise (total energy conservation), and,
second, by effectively performing a weak measurement, since the
emitted photon carries away some information about the electric field  
(encoded in its energy, for example -- see the discussion below).
Comparing the energy shift in Eqs.~(\ref{stat}) and (\ref{polar}) with
the squeezing parameter in Eq.~(\ref{xi}), we may estimate the total
noise energy per cycle  
\bea
\label{gamma-xi}
E_{\rm noise}=\ord(N\xi)
\,,
\ea
with $N$ being the total number of micro-wave photons inside the
cavity. 
This noise energy is of the same order of magnitude 
(with the explicit pre-factor depending on the details of the geometry 
and the dynamics of $\varepsilon_{\rm eff}$ etc.)
as the energy gain due to the dynamical Casimir effect in each cycle  
-- which already demonstrates one of the main drawbacks of spontaneous
decays.  

The estimate of the phase damping rate $\Gamma$ generated by
spontaneous decays is rather complicated. 
To this end, it is necessary to study the entanglement between the
micro-wave mode $\hat a^\dag$, $\hat a$ under consideration and the
reservoir, cf.~\cite{nielsen}.
One indicator is the amount of information about the quantum state
of the micro-wave field (in particular the particle number) leaking
out -- which can be used as a lower bound for $\Gamma_{\rm decay}$.
Ergo, the question is: how much can we learn about the number of
micro-wave photons inside the cavity by looking at the (optical)
photons emitted during the spontaneous decays?
Since the spontaneous decays have to be repeated with a frequency of
order $\Omega$ in order to achieve resonance \cite{exp}, the
line-width will also be of order $\Omega$.
Let $M$ denote the number of absorbed Laser photons -- and hence
excited three-level systems.
Of course, the same number $M$ of (optical) photons must be emitted by
(approximately independent) spontaneous decays in each cycle and hence
the variance of the total emitted energy is $\ord(\sqrt{M}\Omega)$.
This quantity can be contrasted with the accuracy necessary for
measuring the total number $N$ of micro-wave photons inside the
cavity (by looking at the optical photons emitted). 
In the presence of $N$ micro-wave photons with an energy of order
$\Omega$ (resonance condition), the total energy difference in
Eqs.~(\ref{stat}) and (\ref{polar}) is of order $N\xi$, 
cf.~Eqs.~(\ref{xi}) and (\ref{volume}). 
The ratio of these quantities $N\xi/(\sqrt{M}\Omega)$ determines the
information gained in each cycle (i.e., spontaneous decay).
As a result, this special type of decoherence becomes important if the  
number of micro-wave photons $N$ exceeds the threshold set by the
number $M$ of absorbed Laser photons 
\bea
\label{Gamma-xi}
N\geq\ord(\sqrt{M}\,\Omega/\xi)
\;\to\;
\Gamma_{\rm decay}\geq\ord(\Omega)\gg\xi
\,.
\ea
However, as demonstrated in Section \ref{Reservoir}, there are far
more possibilities for inducing phase damping.
For example, if the electrons are pumped (by the Laser) into the
conducting band (in contrast to the three-level system discussed above),  
many more degrees of freedom are available for carrying away energy
and information. 

The undesirable consequences of spontaneous decays can be avoided by
inducing stimulated transitions which are much faster than the
spontaneous decay rate -- 
for stimulated transitions, the emitted photons have the same quantum
numbers as the incident light and thus do not carry away energy or
information.  
E.g., for a Laser beam with a constant intensity $\Omega_{\rm R}=\rm const$,
we obtain the well-known Rabi oscillation
\bea
\label{rabi}
\psi_b(t)=\cos(\Omega_{\rm R} t)
\,.
\ea
Rabi oscillations (i.e., stimulated controlled transitions  
$a \leftrightarrow b$ instead of a spontaneous decay $b,c \to a$)
maintain coherence -- e.g., they are used to manipulate QuBits
\cite{nielsen}. 

\section{Conclusions}

The objective was to study the impact of decoherence on the dynamical
Casimir effect.
For the pure decay channel (amplitude damping), the dissipation of
energy results in a subtractive reduction of the characteristic
exponent, i.e., the number of particles only grows exponentially if
their creation (via squeezing) is faster than their decay 
(given by the quality factor of the cavity), cf.~\cite{losses}.  
However, even without energy loss, e.g., for pure phase damping, the
dynamical Casimir effect is diminished by decoherence -- 
but in a different way.
For pure phase damping, the number of created particles always
grows exponentially, but with a reduced exponent.
Furthermore, the pure quantum vacuum contribution 
(i.e., the original dynamical Casimir effect) would only yield a
linear increase of $\langle\hat n\rangle$; the (reduced) exponential
growth is caused by classical resonance amplifying already present
particles. 
For an experimental verification of the dynamical Casimir effect, a
large enough quality factor of the cavity is necessary, but   
{\em not sufficient}, the quantum coherence time is also a very 
important quantity. 

In order to apply the above results to a concrete example, we
considered a three-level system as a simple microscopic model allowing 
the coherent control of $\varepsilon_{\rm eff}(t)$ via Laser
illumination (of a semi-conductor, for example).
As long as the Laser beam can be treated as a classical external field 
(stimulated emission only), the evolution is unitary, i.e., without
decoherence (under the assumptions made).
Spontaneous decays, on the other hand, inevitably generate
decoherence: firstly, by dissipating the polarisation energy of the
medium in Eq.~(\ref{polar}); and, secondly, by allowing information to
leak out -- which effectively corresponds to weak measurements.
Although the explicit decoherence rate depends on the concrete 
realisation, spontaneous decays are therefore not desirable for a
controlled experimental verification of the dynamical Casimir effect.

As one possible solution for this problem, one could use Rabi
oscillations which are faster than the rate of spontaneous decays and  
automatically generate a harmonic dependence of 
$\varepsilon_{\rm eff}(t)$, cf.~Eq.~(\ref{rabi}), 
maintaining coherence. 
Another advantage of Rabi oscillations lies in the fact that they do
not deposit energy into the material, i.e., there is (ideally) no
heating -- avoiding unwanted excitations which generate decoherence. 

If one departs from the three-level system consisting of localised 
electrons and excites electrons in the conducting band, similar
difficulties arise.  
Apart from the aforementioned problems owing to spontaneous decays, a
finite conductivity always entails additional noise and decoherence
according to the fluctuation-dissipation theorem.
Furthermore, one should be very careful not to excite degrees of
freedom generating a non-linearity which induces a coupling to higher 
harmonics.
E.g., via parametric down-conversion, such a coupling might overcome
the frequency separation between the Laser and the micro-wave photons
-- and thereby completely swamp the desired signal.

\ack

R.~S.~acknowledges valuable discussions during the 
{\em International Workshop on the Dynamical Casimir Effect} 
in Padova/Italy 2004 (see \cite{exp}) as well as financial support by
the Emmy-Noether Programme of the German Research Foundation (DFG)
under grant No.~SCHU 1557/1-1 and by the Humboldt foundation.  

\newpage

\section*{References}


\begin{thebibliography}{10}

\bibitem{casimir}
Casimir HBG 1948 {\em Kon.\ Ned.\ Akad.\ Wetensch.\ Proc.} {\bf 51} 793

\bibitem{moore}
Moore GT 1970 \JMP {\bf 11} 2679

\bibitem{micha}
Uhlmann M \etal 2004 \PRL {\bf 93} 193601

\bibitem{exp}
Braggio C \etal 2004 \RSI {\bf 11} 4967;\\
(see also
{\tt quant-ph/0411085} 
and 
{\tt http://www.pd.infn.it/casimir/})

\bibitem{losses}
Dodonov VV 1998 \PR A {\bf 58} 4147

\bibitem{squeezing}
Dodonov VV 2002 \JOB {\bf 4} R1-R33

\bibitem{mandel}
Mandel L, Wolf E 1995 
{\em Optical Coherence and Quantum Optics}\\ 
(Cambridge: Cambridge University Press)

\bibitem{scully}
Scully MO, Zubairy MS 1997
{\em Quantum Optics}\\
(Cambridge: Cambridge University Press)

\bibitem{nielsen}
Nielsen MA, Chuang IL 2000 
{\em Quantum Computation and Quantum Information}\\
(Cambridge: Cambridge University Press)

\bibitem{via}
Dalvit DAR, Neto PAM 2000 \PRL {\bf 84} 798

\end{thebibliography}
\end{document}